# Server, server in the cloud. Who is the fairest in the crowd?

**Abstract**

This paper follows the recent history of automated beauty competitions to discuss how machine learning techniques, in particular neural networks, alter the way attractiveness is handled and how this impacts the cultural landscape. We describe experiments performed to probe the behavior of two different convolutional neural network architectures in the classification of facial attractiveness in a large database of celebrity faces. As opposed to other easily definable facial features, attractiveness is difficult to detect robustly even for the best classification systems. Based on the observations from these experiments, we discuss several approaches to detect factors that come into play when a machine evaluates human features, and how bias can occur not only in data selection but in network architectures; in multiple forms on multiple levels throughout the process. The overall goal is to map out with mixed methods a novel condition: slippages produced by platform level machine learning systems that make judgements in domains considered dependent on high level human intelligence.

## Introduction

In 2016 Youth Laboratories launched what they called the first beauty contest evaluated by robots; a beauty contest to end all beauty contests.

That was the plan.

What the project, aptly dubbed *Beauty.AI,* did achieve was almost instantaneous notoriety [WIRED]. The 44 male and female winners -from the over 7'0000 entrants who submitted selfies through the organization's app- were mostly white and young. Social media smelled blood in the racist outcome of the robot contest and *Beauty.AI* went offline.

While the subsequent story of how the savvy internet entrepreneurs behind *Beauty.AI* found a clever way to relaunch their endeavor with a slightly socially sensitive version of the original project dubbed *Diversity.AI* might interest some readers, this text will not dwell on that story nor the obvious scandal the event produced, but rather focus on the long tail of events that allowed it to unfold in the first place.

## Beauty and the machine

*Beauty.AI* is but one of several recent business minded attempts at using computers to evaluate human beauty. Agnostic to the historical contingencies of beauty, the appreciation of Rubenesque roundness or other alternatives to the current paradigm of beauty, are not part of their agenda.  In 2016, the startup *Blinq* experimented with an attractiveness detector within an existing online dating site to such fanfare that the creators decided to launch it as a dedicated service: *HowHot.io*. This service did away with any political correctness and focused on the meat: anyone offering selfies to the developers could have their sexual attractiveness - and only that - automatically rated by the system [TechCrunch].

*Beauty.AI*, *Blinq*, and *HowHot* are indicative of several threads that have been pulling beauty and attractiveness onto the center stage of the computer age. Entrepreneurial startups and university spin offs are at the forefront of making commercial use of the synergies fueled by phones, algorithms and desire. The *Beauty.AI* contest was launched by the startup Youth Labs that specializes in aging research [Silico]. The software which evaluated the *Beauty.AI* image submissions was based in part on the company's ongoing research into wrinkle detection and skin condition monitoring [RYNKL]. Likewise the *Blinq* site was created by enterprising graduate students from the Computer Vision Lab of ETHZ; a lab that specializes in image analysis of medical images and scene understanding [ETHZ][1].

The obsession with automated beauty detection is not limited to ambitious startups. Even recent academically oriented inquiries into attractiveness justify the significance of their research with its potential relevance to the booming beauty industry. A group of researchers from Tsinghua and Hong Kong Polytechnic Universities, for example, published a novel facial geometry based hypothesis on facial beauty perception and noted the project's usefulness for cosmetic "surgery plans, online dating recommendations and photo retouching" [Chen2014].

The history of algorithmic evaluation of facial beauty is over a decade old. And from the start, there was a functional underpinning to the inquiry. If one were to use Kant's categories of the beautiful to describe the approach of algorithmic evaluation, one could say that it was not a freely imaginable universal beauty, detected by judgement without consensus, but a type of beauty 'attached to' (some form of) intention[2], and endowed with regularity and symmetry; in short a utilitarian version of beauty detectable with geometric means.

---

[1] At the time of this writing, neither *Blinq* nor *Beauty.AI* are offering their beauty evaluation services.

[2] " An einem Dinge, das nur durch eine Absicht möglich ist, einem Gebäude, selbst einem Tier, muss die Regelmässigkeit, die in der Symmetrie besteht, die Einheit der Anschauung ausdrücken, welchen den Begriff des Zweckes begleitet, und gehört mit zum Erkenntnis." [Kant 1790].

In 2006, a research group from Tel-Aviv University laid the groundwork for machinic evaluation of facial attractiveness [Eisenthal2006]. Beginning with a review of pre-computational based studies of the assessment of facial beauty across cultures and ethnicities, Eisenthal's group evaluated detectable features (such as small noses, prominent cheekbones and facial proportions) hoping to find beauty criteria that are not in the eyes of the beholders. Failing to find a generalizable solution to the challenge based on geometric properties, the researchers turned to "using the images themselves" with attractiveness scores determined by human raters in a novel machine learning based beauty detection approach. While the approach set the stage for machine learning beauty evaluation, the experiments performed were hampered by the limited datasets collected by the research team: two datasets of 92 images each[3] which were evaluated with traditional statistical classification techniques[4]. In hindsight, it is not surprising that Eisenthal's group was not able to achieve anything over 65% accuracy in beauty prediction.

The more recent resurgence of interest in machinic beauty detection is enabled by at least three different vectors. First, the social media fueled obsession with youth culture supported with endless streams of increasingly high resolution images enabled by platforms such as Flicker, Instagram and others. Second, supporting the first point, the ease with which images can be created and distributed via ubiquitous and cheap mobile phones. Third, advances in image processing techniques that allow one to extract information from images robustly and rapidly. This third item in particular has been enabled by the popularization of neural network architectures designed specifically for image processing at scale.

## Beauty in engineering benchmarks

In computer science the collection and curation of datasets is far less scrutinized than the algorithms that operate on them. In fact, the provenance of data can remain obscured while data is used, and reused. An early case in point is the infamous test image used in computer vision research, Lenna (figure 1). The test image is a picture of Lena Söderberg, cropped from the centerfold of the November 1972 issue of *Playboy* magazine [WikiLenna]. Lenna became one of the most widely used benchmark tests in early computer vision [Hutchison2001]. Lenna does double duty. She offers challenges to algorithms seeking to prove their worth in the compression of varied image features such as shading and texture, and Lenna offers a pleasant view to the mostly male computer programmers who spent endless hours assessing the performance of their codes on this benchmark. The male gaze and determination converge in the handling of Lenna as master-image-muse. The affair however did not last. Lenna no longer suffices as a benchmark, and recent alternate benchmarks for computer vision tests have included examples of reverse sexism with the male model Fabio Lanzoni (figure 2), this time carefully scrutinized by female computer vision researchers [Needell2013].

Lenna and Fabio will not help us find beauty in the new regime of deep learning and big data; individual images no longer play the same role in big data regimes that they did in benchmark image investigations. Salient information is gleaned from patterns detected across not one but a collection of images. Likewise any attempt to scrutinize the influence of the designers and programmers must be sought with a different

[3] Despite the small sample size, the researchers created on online system to generalize the rating process. They also used a large number of observers relative to the size of the image base: 28 male and female raters for the first image set alone.

[4] Principle component analysis to reduce the dimensionality of the image space and K nearest neighbors to cluster the images based on attractiveness scores

approach. In order to better understand the various dependencies between engineering design, human desires and unquestioned biases in deep learning technologies that seek to detect beauty and attractiveness, a detour into a specific computing technology, namely convolutional neural networks, is necessary.

## Beauty and the machine 2.0

Eisenthal's group set the stage for subsequent inquiries into algorithmic beauty detection. Researchers revisiting the beauty detection make use of two significant additions to Eisenthal's toolbox. One, large data sets and two, new approaches of enabling computers to learn from data.

Of all the learning approaches that have been developed over the past decades, convolutional neural networks (CNNs) are the most significant class of techniques. CNNs constitute the current best practice in image analysis, outperforming previous approaches in automated image analysis [LeCun1995]. We will concentrate our discussion on this category of supervised learning systems.

## Convolutional neural networks

CNNs, like neural networks, are inspired by biological networks and made up of nodes selectively modeled after neurons. Each neuron-node receives several inputs, takes a weighted sum over them, passes it through an activation module and responds with an output. As opposed to standard neural networks, CNNs use images as inputs. As the preprocessing steps churn the image information, increasingly high-order features are extracted from an input image. At the last stages, these features are fed into a fully connected neural network which in turn produces a classification. Figure 3 depicts a diagram of a standard CNN architecture.

Convolution and down-sampling are key to reducing the abundance of image information to a smaller but functionally equivalent set at the first stages of processing. Convolution consists of moving a filter (a small matrix of values) across the entire image, multiplying the elements of the filter with the image pixels it overlaps with, and replacing the pixel at the center of the filter with the product of this operation. Max pooling, a particular form of down-sampling, is then applied to this output. Max pooling reduces the dimensionality of a data set (reducing its size). Together these operations create an abstraction of the original image and prevent the subsequent neural net from over-fitting to the details of the original image, helping the network to learn the general, not the specific, features contained within an image[5].

CNNs have typically been used in a standard structure; stacked convolutional layers followed by one or more fully-connected layers. Variations of the CNN design have been used on benchmark image collections (such as MNIST[6] and CIFAR[7]) with very good results [Szegedy2014], [Krizhevsky2012]. Performance improvements have often come from increasing depth (number of levels) and width (number of units at each level) of the standard structure. Size matters; there is a trade-off between network size and training data size. The larger the network, the larger the use of computational resources and the higher the likelihood of overfitting [Szegedy2014], i.e. creating a model that responds well to a given set of data but does not generalize to fit to other data sets. Indeed, flooding a low-complexity architecture with too many examples can over-train (‘overfit’) a network and make it simply mimic the data as opposed to being able to generalize across the data examples.

---

[5] The ability of a network to generalize is dependent on many other factors not discussed here (including dropout rate).
[6] http://yann.lecun.com/exdb/mnist/
[7] https://www.cs.toronto.edu/~kriz/cifar.html

More often than not, however, data sparsity is more of a problem than data abundance. Moreover, not all data is good data, and so researchers have crafted methods to artificially augment image data sets by copying images and then applying modifications to the copies - such as minor rotations - that introduce small changes to the ‘new’ images but leave the content of the original image largely untouched [Simard2003]. Related to such data augmentation strategies, some advanced CNN architectures are ‘pre-trained’, i.e. exposed to a standard dataset before they are fine tuned to a specific task. More on this below.

**Beauty and the network architecture**

In order to better understand how beauty is represented in CNNs we have created a series of experiments. Our approach seeks to understand both the role of the classification architectures in the detection of attractiveness and beauty as well as the role of the data and data collection applied to the training of the classification architectures.

Our departure point is a recently released database of celebrity photos, *CelebA*, compiled by the Chinese University of Hong Kong containing 202,599 face images and 40 binary attributes/features [Liu2015]. Labeled datasets of this size are typically available to enterprises and not the general public. As such it serves as a good entry into both the evaluation of architectures as well as the questioning of data collection practices.

Our first experiment (E1) used a fully connected standard configuration CNN (see figure 3 for details). We implemented this network in the open source *TensorFlow* environment. Here is a summary of the results produced by this network:

Table 1: Select results from experiment #1 (Vanilla CNN)

| feature # | feature name | accuracy | # training imgs | # test imgs | img size |
|---|---|---|---|---|---|
| 2 | attractive | 50.0% | 5000 | 5000 | 32x32 |
| 2 | attractive | 51.2% | 25000 | 25000 | 32x32 |
| 2 | attractive | 51.5% | 5000 | 5000 | 64x64 |
| 4 | bald | 97.8% | 5000 | 5000 | 32x32 |
| 13 | chubby | 94.3% | 5000 | 5000 | 32x32 |
| 15 | eyeglasses | 93.3% | 5000 | 5000 | 32x32 |
| 16 | goatee | 93.4% | 5000 | 5000 | 32x32 |
| 26 | pale skin | 95.9% | 5000 | 5000 | 32x32 |
| 29 | rosy cheeks | 93.3% | 5000 | 5000 | 32x32 |
| 39 | young | 78.1% | 5000 | 5000 | 32x32 |

While E1 is a standard CNN configuration used in image classification, it is a rather simple network. Nonetheless, E1 was able to robustly detect 24 of the 40 features with an accuracy above 80%, and 12 features with an accuracy above 90% when fed with 50k images[8]. Of interest to our inquiry is the fact that the feature ‘attractive’ remained with a maximum achieved accuracy of 51% essentially unlearnable to this network, while other features such as ‘young’ or ‘rosy cheeks’ were learned robustly. Indeed, the feature ‘attractive’ remained unlearnable to this network architecture despite several variations to this CNN architecture.

In order to understand the uneven performance of this network, we tested the same image set on a second CNN network. In this second experiment (E2) we used the substantially more complex *Inception Model* (figures 4 and 5). Inception architectures are larger than standard CNNs and offer two important advantages, namely the network-in-network approach and dimensionality reduction. The network-in-network approach applied to CNNs means that convolutional building blocks are sequentially combined and integrated into the computational pipeline. Dimensionality reduction alternates convolution kernel sizes to reduce the overall number of required computations. Importantly, the Inception model we made use of contains over a million images with 1000 labeled categories [ImageNet], so any new CNN architecture using this framework already has a general idea of a large set of everyday categories [Szegedy2014]. The significance of the ‘pre-learning’ is discussed from a technical perspective in the appendix (textbox1) and will be revisited conceptually later in this text.

The table below summarizes our results. The feature ‘attractive’ that the first network could not detect, can be found in the second network with close to 80% accuracy. Far from perfect, this is a significant improvement over the performance of the first network.

Table 2: Select results from experiment #2 (Inception Model with CNN and pre-training)

| feature # | feature name | accuracy | # training img | # test imgs | img size |
|---|---|---|---|---|---|
| 2 | attractive | 74.7% | 1500 | 500 | 199 * 199 |
| 2 | attractive | 76.3% | 15000 | 5000 | 199 * 199 |
| 13 | chubby | 79.5% | 1500 | 500 | 199 * 199 |
| 13 | chubby | 83.7% | 15000 | 5000 | 199 * 199 |
| 15 | eyeglasses | 90.6% | 1500 | 500 | 199 * 199 |
| 15 | eyeglasses | 93.5% | 15000 | 5000 | 199 * 199 |
| 16 | goatee | 83.7% | 1500 | 500 | 199 * 199 |
| 16 | goatee | 89.3% | 15000 | 5000 | 199 * 199 |
| 39 | young | 75.8% | 1500 | 500 | 199 * 199 |
| 39 | young | 83.3% | 15000 | 5000 | 199 * 199 |

[8] The network’s performance did not notably increase with the full 200k image set.

The research team that collected the *CelebA* image set has reported even better classification results. Their architecture has been able to learn the feature ‘attractive’ with accuracies between 78 and 81% [Liu2015]. However, that team’s approach included a preprocessing step specific to the *CelebA* image set with finely tuned and cascaded CNNs where the first network evaluates face localization and the subsequent network then extracts face features [Liu2015].

These differences are interesting in several regards. One of the advantages of neural networks is that they are said to require less technical expertise than other data evaluation techniques such as traditional statistical methods. This is one important reason for their current popularity. Several off-the-shelf CNNs produce surprisingly good results on select image sets (such as the older benchmark MNIST). Some researchers have formulated best practices for CNNs for visual tasks that make use of only a few of the many features network designers have at their disposal [Simard2003]. Indeed the recent release of several high level libraries with neural net architectures such as *Tensorflow, Caffe, Torch* and *Theano* [Bahrampour2015], allow easy access to basic architectures, and leave the many details unexplained for the novice to consider. As the results from our vanilla CNN show, even a comparatively simple architecture can robustly detect several visually salient features (such as the presence of eyeglasses).

From our experiments we conclude two things. Firstly - and this part will surely be evident to CNN experts - is that details of the architecture greatly matter. Secondly - and this is more surprising - the architecture details don’t matter across all detectable features uniformly. Indeed, some of the more visually salient features were learned with higher accuracies in the simpler of the two architectures. In one sense, the feature ‘attractive’ we have been discussing operates as a detector of the significance of CNN architecture choice. Thirdly, we see that none of the approaches are able to detect the feature ‘attractive’ as robustly as more physical features such as sideburns or facial geometry. As such, this feature represents an unusual case.

While it is conceivable that a yet more advanced future CNN will deliver yet better results for this category, we want to attempt to uncover why the current approaches are so unsuccessful, and then reflect the results from that attempt back onto the discussion of networks.

Attractiveness, even when reduced to what is perceivable in a face, is a complex human judgement generated in a mix of objective, subjective and cultural dimensions. Importantly, we want to discuss now not only how CNN architectures matters, but include other factors that computer professionals usually do not consider equally relevant to the classification process. We will unpack this problem with several different tools, some computation and others observational.

## Beauty and the data

If one wants to understand the processing of beauty in the machine, one has to look specifically at two distinct parts of the system: the decision mechanism and the data. Neural networks have in the past been assailed as ‘nontransparent’ [Yosinski2015] because they do not allow a human user to easily ‘see’ intermediate results nor understand how they logically lead to the outcome. As opposed to an algorithm operating with symbolic logic, neural-inspired architectures store all intermediate results in the form of multi-dimensional matrices, the weights of which encode the current state of the system. 'Looking at' these values give human beings no intuitive insight of what the system is up to. In response to this lack of transparency, several researchers have proposed methods by which one can make the internals of neural networks more understandable.

In our case, we want to understand how and if 'attractiveness' differs from other features in terms of its learnability by a machine. One way to do begin to understand the relationship between this feature and its 39 co-features of this particular dataset is to see how the collection of all features depend on each other with a co-occurrence map (figure 6). This plot of the co-occurrence relationship between the 40 labeled features of the *CelebA* image set shows one possible reason why 'attractive' is a special case. While features 'chubby' and 'sideburns', for example, appear jointly frequently, the feature 'attractive' does not correlate with any of the other features in an apparent manner. It appears that attraction, as defined by the group who produced the labels, is by and large an independent feature; i.e. it does not appear together with other labels in a structured way. However, the co-occurrence map does not give us any information as to *why* the feature 'attractive' would be any more challenging to detect than the others.

For additional insight into the complexity of the 'attractiveness' feature we created a decision tree classifier that uses the 39 co-features for each image as the independent variables to classify the image as 'attractive' or 'not attractive' (see textbox 2). Figure 7 shows a depth 5 tree trained on a subset of the *CelebA* data set. The decision tree confirms the observations from the co-occurrence view, namely that the label 'attractiveness' in this data is not clearly associated with the other labeled features.

**Beauty and confusion**

As opposed to the computational network dynamics, the data selection and labeling processes that precede classification have received less structured scrutiny. To engineers and consumers, data are by and large assumed as given, even though they are always already "cooked" [Bowker2013]. Likewise, data as images are always already prepared in one way or another. Simply by selecting 'this image' as opposed to 'that image' to represent a given feature, category boundaries are created long before machines operate on them. By looking more closely at the data within the *CelebA* collection we will try to show not only that data selection and labelling are integral parts of the classification apparatus, but that the very collection background in which these actions occur also materially matter.

According to the Hong Kong research group, *CelebA* dataset was annotated by a group of 50 paid male and female participants, aged 20 to 30, and recruited from mainland China during a 3 month development phase[9]. As the title of the collection suggests, the images depict famous persons - celebrities - from around the world. The images of these celebrities were collected from various online sources where copyright allowed[10]. Once the collection effort was complete, however, the actual work began. Labelling 200'000 images is a time-consuming task. Given the tight three month deadline, each of the 50 workers must have labeled about 70 images/hour, each with 40 features, resulting in about 50 unique decisions a minute, assuming an eight hour/day workload (figure 8). That would be a substantial amount of work performed seven days a week continuously. It is not surprising then that a few errors might occur.

[9] Personal communication with Ziwei Liu, approximately one dozen email messages between June 9th and July 8th 2017. We were not able to receive access to the actual source code used in this group's experiments.

[10] http://mmlab.ie.cuhk.edu.hk/projects/CelebA.html

One obvious disadvantage of big data sets is that is becomes increasingly difficult to simply view the image data. While this condition might not matter for some labelling tasks, it does matter where categories are as complex as the illusive feature ‘attractive’. Indeed, simply looking at the data set is revealing. Figure 9 shows images labeled as ‘attractive’ and figure 10 shows different images labeled as ‘not attractive’. It is not difficult to see that several of these images do not seem to fit in their defined categories. It is then no surprise that a sizeable confusion matrix - a collection of false positives and false negatives - can be established with our Inception model CNN (figures 11 and 12). In some cases the errors are far from subtle. Even the feature eyeglasses has mislabeled entries in the dataset (figure 13).

Anthropologists have long pointed out the complexity of defining facial attractiveness. Krzysztof Kościński’s in depth overview of the literature on facial attraction describes how situational factors impact the judgment of facial attractiveness. The age group of the participants, for example, is assumed to play a significant role in the evaluation of attractiveness. The older a person is, the older the faces he/she prefers [Kościński2008, p80]. Furthermore, the physiological state of testers can impact the decisions at evaluation time. In the fertile phase of a woman’s cycle (preovulatory period) women evaluating male facial attractiveness display stronger preference for masculinized male faces [Kościński2008 p81], and people in stable bonds tend to give lower assessments to unfamiliar male faces [Kościński2008 p83] while even a moderate consumption of alcohol makes faces of the opposite sex seem more attractive [Kościński2008 p83]. Likewise, there is substantial evidence for the fact that attraction to a face is attraction to a previously seen face [Kościński2008 p85]. Global Celebrities are strongly western biased in demeanor and appearance, making the evaluation of the dataset by a group of young Chinese workers, poorly remunerated and working long hours, an awkward exercise in cross-cultural exchange; a commentary maybe on the propagation of western attributes as a function of exposure to (ubiquitous) western beauty ideals that has been clearly observed in other contexts. [Kościński2008 p. 86].

Facial reading has different histories in different cultures, including *mien shiang* in China, *gwansang* in Korea, *kao no dokusho* in Japan [Hutchinson2017] [Mar1974], and it is not clear to what extent the cultural thread of mien shiang, traditional Chinese face reading, lingers in the evaluation of the participants in this exercise. The influence of these age-old traditions are far from academic curiosities. Plastic surgeons operating in Hong Kong report that they are often asked by patients to perform surgery that will ‘alter their fate’, and some surgeons self-describe their practice as providing a service wherein “Western surgical practices must often address Eastern aspirations” [Wong2010], allowing patients who can afford the price of a life without a “poor man’s chin” (figure 14) to change their fate, for example. The point of these digressions - and speculations - is to shed light on the under-observed aspect of CNN activity, that of data creation and curation, and to declare these activities bona fide sites for investigation; investigations for which engineering disciplines are ill-equipped.

Supervised machine learning techniques such as CNNs require large amounts of data, and somewhere in that data collection process there are real people with cultural baggage making practical decisions. While large image sets are easy to compile, large annotated datasets are cumbersome and expensive to create. This problem only grows with the tendency of CNNs to perform better with large datasets and exerts pressure on researchers to create data aggressively with potential sacrifice to details. The manual image labelling process is hardly a well-controlled process, and is itself a source of unknowns. In the case of the *CelebA* dataset, the cleanly arranged binary labels suggest crisp decision boundaries that belie the complexity underlying the hidden annotation process. For this inquiry, the binarization of the 'attractiveness' feature is of interest; it in no way does justice to the complexity of influences underlying the evaluation, that - in the case of 'attractiveness' occurs almost subliminally [Olson2005] as humans tend to judge faces with instinctual rapidness. Other objects such as abstract paintings [Duckworth2002] or animals [Halberstadt2003], have been reported to require longer scrutiny before they can be deemed attractive or not.

**Computational ageism**

While the question of what constitutes beauty and attractiveness has been pondered for millennia, recent research suggests that both biological as well as social and cultural factors play operational roles in its formation [Dion1972], [Rhodes1998], [Etcoff1994], [Griffin2006]. Furthermore, the complexity of beauty and attractiveness that even social media pundits recognize [Slate2016] has been confirmed by psychologists describing the multimodal nature of attractiveness more formally [Groyecka2017].

Across all these inquiries and including recent surveys of beauty perception in non-western cultures [Coetze2012], youth ranks as one of the most singularly significant components across cultures. Whether this is due to enduring biological constraints, including "mate value based on perceivable ability to procreate" [Kościński2008] or economic and social constructions seems undecided at this point, and we are certainly in no position to contribute to that discussion; and it is not the point of this paper. However, attractiveness is far less objective than age, and age can be assessed much more precisely with computational methods [Rothe2015], [Rothe2016]. Indeed, the fact that youth assessing computational methods are by and large CNN based, makes this fact relevant for our investigation[11]. One reason CNN methods are applicable to the problem is that social media make image-age pairings comparatively easy to obtain in large quantities. And because the results are by and large accurate [Rothe2015], they receive more attention than features that are less crisp and harder to evaluate. Because youth is a significant component of beauty and because youth can be more readily assessed computationally, its valency changes to an attainable proxy of the more elusive concept of beauty. And this computationally driven shift then allows that which is calculable to become more common. It changes the discussion and culture of how a networked society experiences beauty and attractiveness and creates the potential for a new form of computationally enabled ageism. As such, the apparatus of big data enabled CNNs alter the way attractiveness is managed in the cultural landscape. This effect may be similar to the way cosmetic surgery can have a lasting effect on beauty standards because it has at its disposal a viable technology, and then responds pragmatically to market demands for new faces which in turn then drive the way beauty is seen and lived with.

**Understanding what algorithms do**

There is growing concern about and interest in understanding how neural networks, including CNNs make decisions. In military robotics for example, neural networks that make real time battlefield decisions are being augmented with logic-based systems whose step by step operations can be understood by human beings and whose structure is more compatible with the formal rules of war and peace conventions [Arkin2012]. Similarly, DARPA has launched a large effort to create Artificial Intelligence systems that are specifically designed to be human-understandable. *Explainable AI*, as this effort is referred to, aims to make systems that human beings can not only understand, but also trust [Samek2017]. Yet another need for understandable systems comes from the fact that human perception is inferior to that of computers even in some socially sensitive areas. Wang & Kosinski recently 'demonstrated' that a deep CNN classifier can distinguish between prepared photos of gay and heterosexual men and women more robustly that human judges [Wang2017], and the backlash from many sides followed promptly [Murphy2017] as these researchers also failed to understand the significance of data curation and uncritically followed the biases produced by the network they deployed. It is no surprise that algorithmic fairness has become a fundable research agenda [Albarghouthi2016].

[11] Even our most rudimentary CNN architecture was able to detect the label 'young' with over 78% accuracy. See Table 1.

Specific to the domain of neural networks for machine vision, researchers have proposed to visualize the activities of select network layers or even individual neurons in a network [Yosinski2015]. While this approach has demonstrated real promise in making the activation events inside of CNNs visible and has synced them with visible expectations of human observers comparing input images with the network activities, it also demonstrates that neural networks simply operate differently that humans do. In some cases, neural networks can be triggered by image elements that human beings pay no attention to. Moreover, 'seeing' the activation (or weight) of a single neuron in the layers of a network is practically meaningless [Tishby2017] as many network internal configurations can deliver the same level of performance. The real action of a neural network reveals itself over time as it transitions from fitting to given data to generalization and reduction of information, the 'compression phase' [Tishby2017]. Using visualization to force neural nodes into a 'graphic mode' and rendering them seemingly 'readable' runs the risk of suggesting simple causes where there are complex relationships;  computational comb-over that does nothing to deeply understand the fundamental features, and potential flaws of decision making within the network.

The problem of non-transparent computer processes has been addressed in other professional fields of informatics with different arguments and motivations. Computer Human Interaction researchers have suggested that understanding of algorithms require people to understand "not only of the process at hand, but of the entire design context and motivations out of which an algorithmic process emerged" [Hamilton2014]. Anthropologists have suggested that the study of algorithms should become a branch of Ethnography or be considered an ethnographic practice [Seaver2013]. While both of these endeavors point, in our view, in the correct direction, they are vague on how one might proceed in any detail.

**When platforms automate human judgement**

It has become apparent that big data practices alter the way research occurs [Williford2012]. The full depth of the fallout of this change is becoming evident to media theorists and historians [Boyd2012], [Carpo2017], suggesting a new need for theories on how to make good use of data-centric computing, not just how to do it efficiently.

Our attempt to show some of the relationships between CNN architectures, data creation and classification results addresses one part of the 'entire design problem' in an experimental manner. In particular our inquiry suggests the need for more focus on the cultural complexity of data provenance and curation underlying big data dependent algorithms.

Yet another critical aspect of neural network based machine learning resides in the class of architectures that incorporate data a priori. As mentioned above, the Inception model approach comes ‘pre-trained’. The motivation for this approach is based on previous research in *transfer learning* [Caruana1998]. When applied to big data dependent neural network architectures, it allows CNNs to be primed with prior experience in the form of exposure to previously collected images. While this added preparation will be useful for some tasks, it is also a tie-in with a very specific representation of world contained within the pre-training data set. Whoever uses Inception models pre-trained with ImageNet, the go-to image collection for machine vision [ImageNet], inherits some of the choices and inherent values of the research team pulling the data together. Bias is not only to be found in the selection of data, but in the selection of classification architectures; this is all the more the case when architectures are optimized for performance and made ‘easy to use’ in pre-trained applications.

More specifically our inquiry circles around a newly significant class of computational-social problems, namely the formation of judgements by computers in areas previously reserved for human beings. As opposed to medical image analysis that used to be routinely performed by medical professionals and is now executed by algorithms at platform level such as the Watson Health Medical Imaging system [WatsonMedical], the problem of human judgement we have started to deal with address not judgment of physical properties such as distance, weights or size, but judgements of taste; judgements within aesthetic categories that do not rely predominantly on expertise but on a combination of hard to formalize factors. While pretty faces can hardly stand in for the sophisticated aesthetic categories and artifacts of human cultural production, they do at least smile at a new class of deficiencies otherwise highly efficient machines will generate.

**Acknowledgements**

Thanks to Andrew Lison for a careful reading of and insightful suggestions to the first version of this text.

## APPENDIX and FIGURES

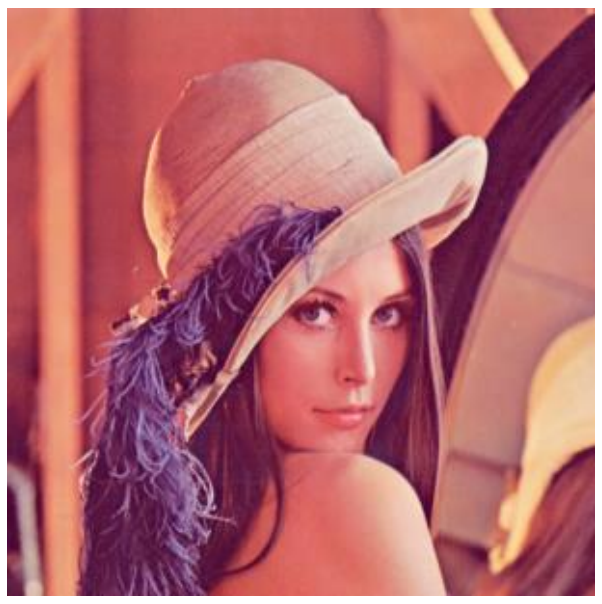

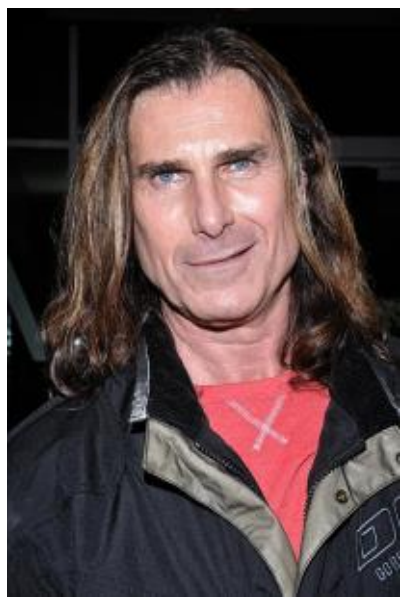

Figure 1. (left) Lena Söderberg 1972, 512x512 scanned section of the Playboy Magazine. November 1972, photograph by Dwight Hooker.
Figure 2. (right), Fabio Lanzoni 2014, photograph by Glenn Francis.

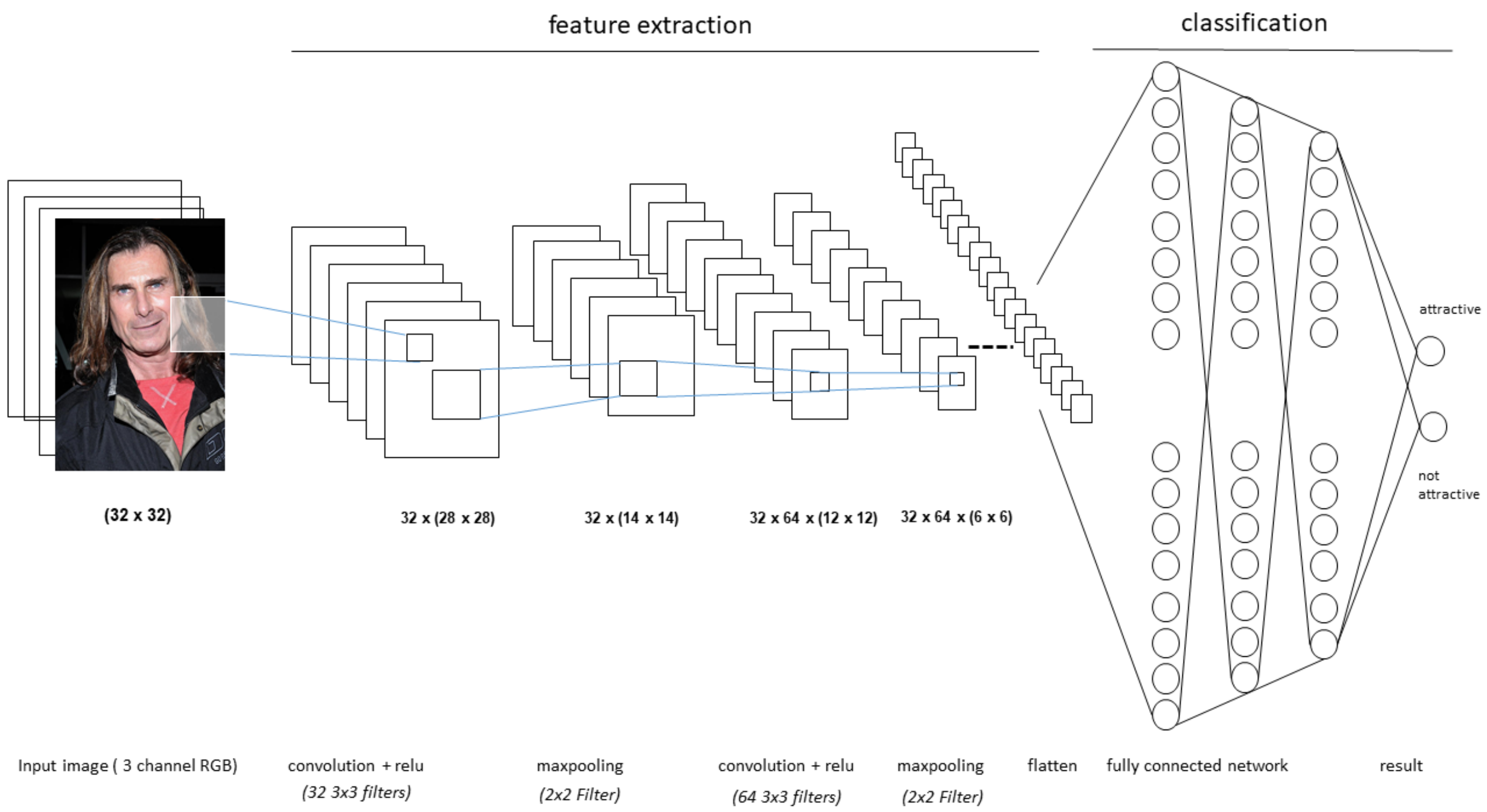

Figure 3.
Standard convolutional neural network architecture used in experiment #1 to classify image data for the label 'attractiveness'. The high dimensionality of image data is reduced by convolution and pooling steps to smaller data footprints that can be fed into a fully connected neural network.

Our experimental network based on this standard model used two convolution and max-pooling steps. The convolution sequence was followed by a fully connected 512 node network with the Rectified Linear Unit (relu) activation function (zero when x < 0 and then linear with slope 1 when x > 0), 50% dropout, followed by a second fully connected network with softmax activation (maps the outputs of each unit between 0 and 1 and normalizes all outputs such that the total sum of the outputs is equal to 1).

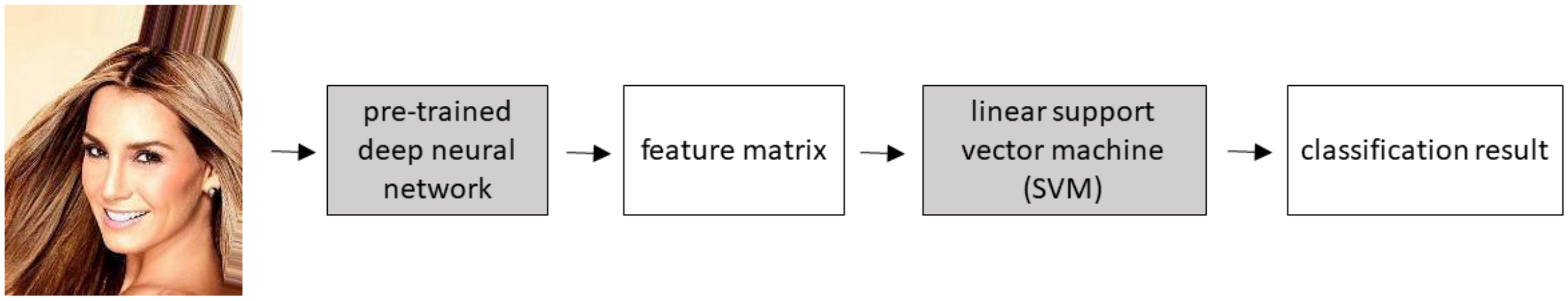

Figure 4.
Simplified schematic of the Inception model approach.

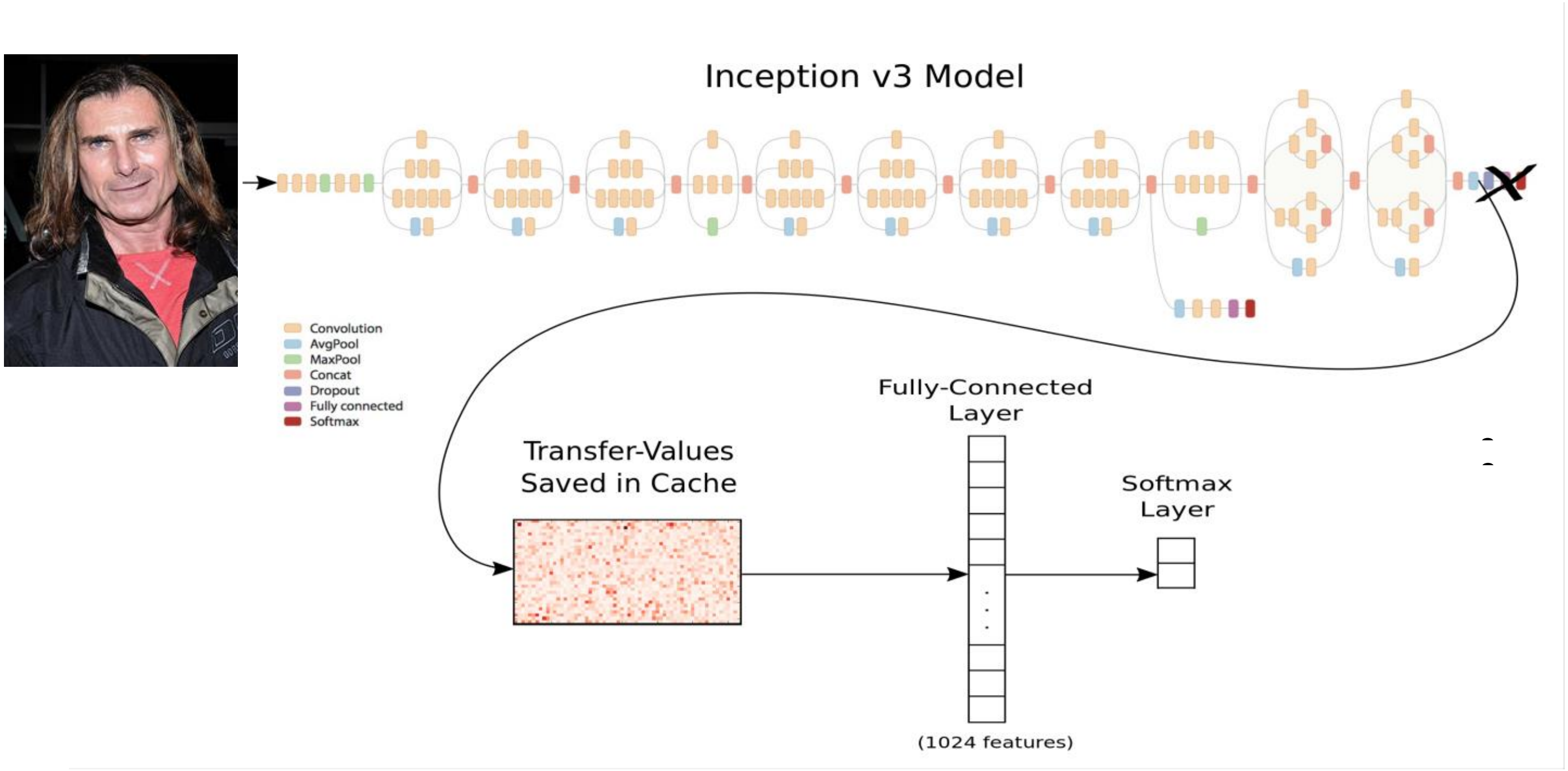

Figure 5.
Schematic of the Inception-3 model created by Google Research [Shlens2016], and modified for our experiment. The Inception-3 model is trained on the ImageNet database and was used in the ImageNet Large Visual Recognition Challenge in 2014. The Inception-3 model is, at the time of writing, one of the most effective models operating on the ImageNet challenge. The Inception-3 architecture consists of a series of factorized convolution modules (that combine 5x5, 3x3 and 1x1 filters) processed at the same input, enabling multi-level feature extraction to generate a matrix of features that is in turn fed into a linear classifier.

**Pre-training networks with transfer learning**

It is becoming a common practice to pre-train a CNN on a very large data set and ‘transfer’ the information gained to different task. Here we describe three approaches used in practice:

a) **CNN as a fixed feature extractor.** If the last fully connected layer of a pre-trained CNN is removed, the remaining network acts as a feature extractor, such that any raw input image will yield a fixed length vector that consist of activations of the hidden layer immediately before the fully connected layer. One can then train a machine learning classifier (for example a linear classifier) on this higher order data for any learning task. Using the output immediately before the classification layer of the CNN is recommended if the images in the current learning task are not similar to the original images used for pre-training. Otherwise, it is better to use the output at a much earlier level within the CNN. This strategy is preferred if the **new training data is small and not similar** to the original data.

b) **Fine-tuning a CNN with new training data**. Another strategy is to continue the training process of the pre-trained CNN, using the new training examples (and the new training classes). This will fine-tune the weights to adjust to the new data. It is recommended that only the later layers are fine-tuned while the earlier layers (corresponding to more generic image features) are not modified, since they are expected to be stable after the pre-training step. This strategy is preferred if the **new training data is large and similar** to the original data.

c) **Pre-trained model weights as initial values**. The last strategy is to train a brand new CNN on the new training data, but initialize the CNN using weights from the pre-trained model. The difference between this and the previous strategy is that here we are fine tuning the entire network. This strategy is recommended when the **new training data set is large but not simila**r to the original data. This approach can lead to faster convergence.

In this study we have the explored the first strategy.

Textbox 1.
Variations on pre-training a neural net (technical discussion)

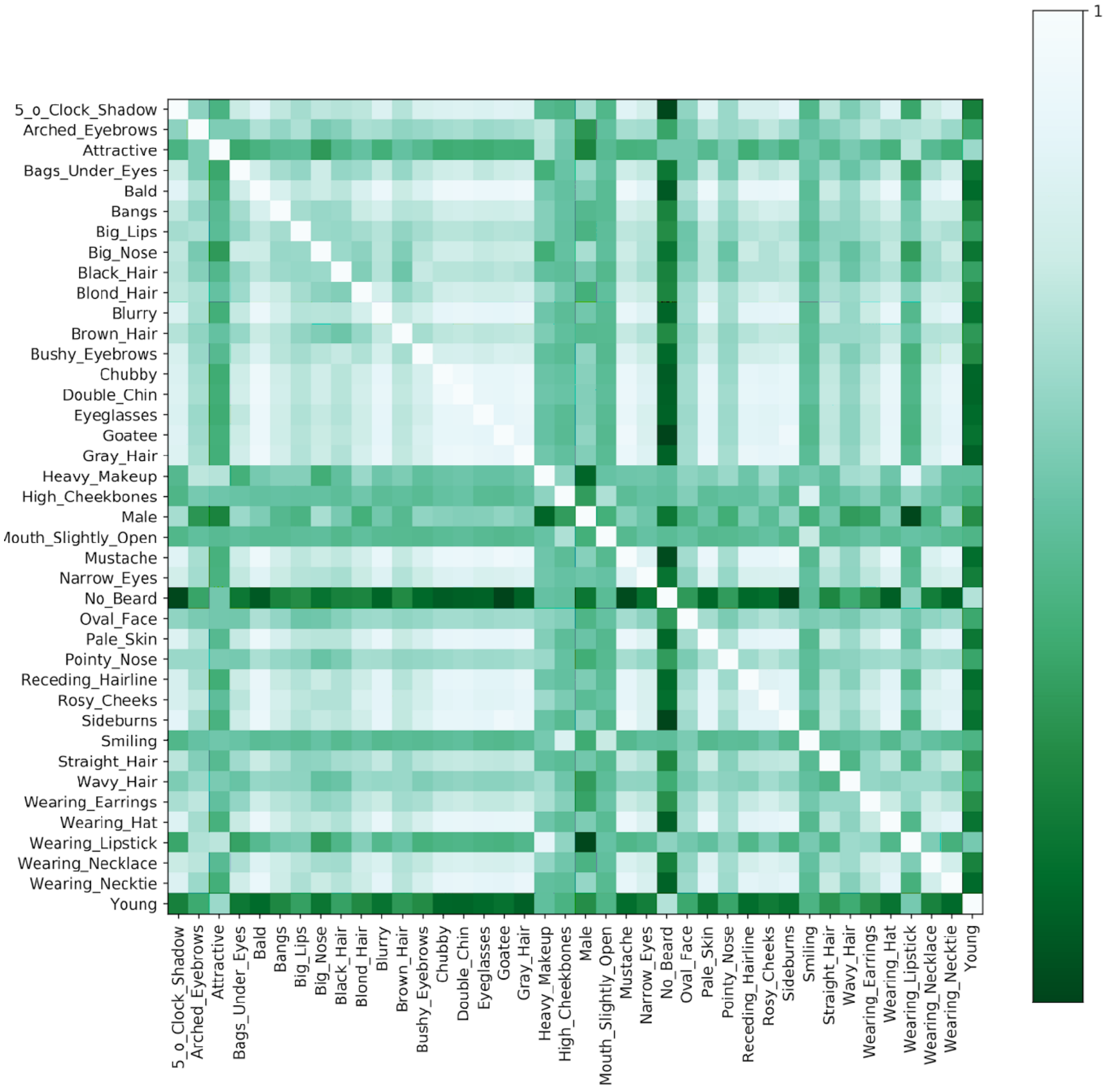

Figure 6.
Co-occurrence map of 40 binary features describing the celebrity photos from the University of Hong Kong's large-scale CelebFaces Attributes Dataset (*CelebA*). White indicates close or high co-occurrence, dark (green) low co-occurrence. The attribute attractive (third from top, third from left) is not strongly associated with any one isolated feature. From this observation we can conclude that the human beings applying the label ‘attractive’ to this data set, determine the feature through a complex process, with internal inconsistencies, thus doing justice to the multidimensionality of the idea of attractiveness.

A **decision tree** is essentially a collection of decision rules, organized as a tree-like structure. Each rule is defined over a subset of the input variables to determine the target class. The learning is done using a greedy search algorithm that seeks to learn rules that can accurately classify the training data examples, but at the same time, are simple enough to guarantee good generalization on unseen test examples. The depth of the tree (length of the rules) is typically used as a user-controllable parameter to influence the simplicity of the learnt tree.

Textbox 2.
Decision trees (technical discussion)

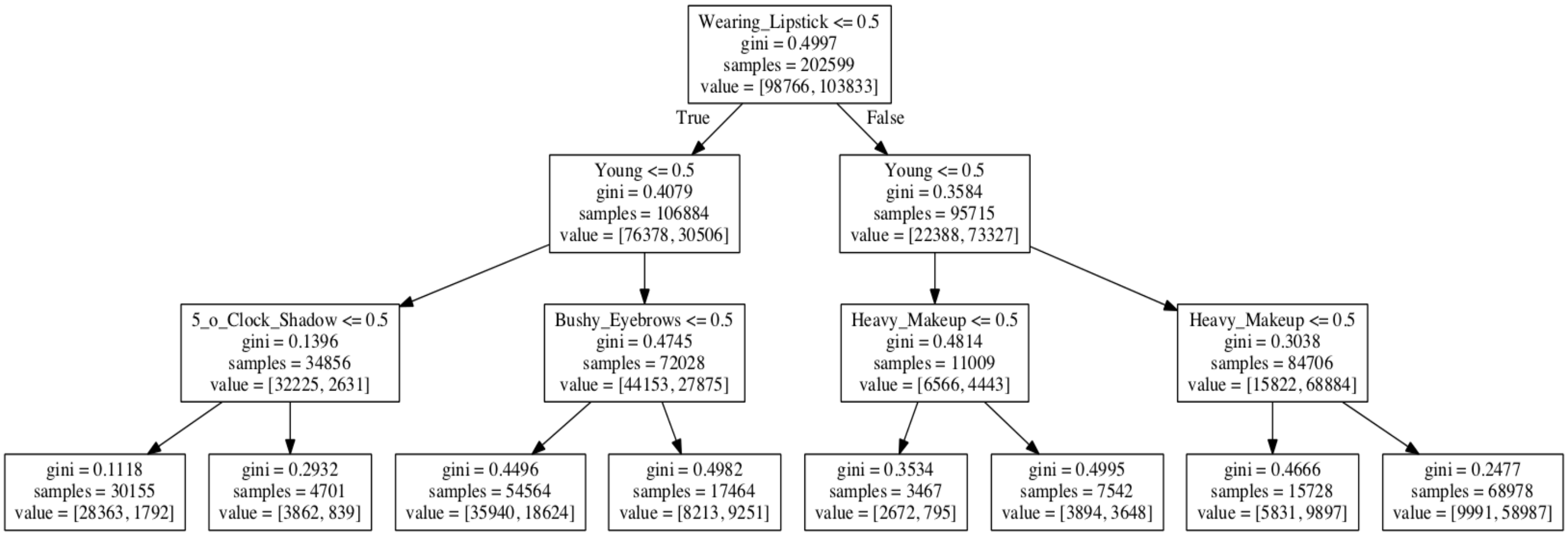

Figure 7.
A depth five tree trained on a subset of the *CelebA* data set.

| | | | |
|---|---|---|---|
| 1 5_o_Clock_Shadow | 11 Blurry | 21 Male | 31 Sideburns |
| 2 Arched_Eyebrows | 12 Brown_Hair | 22 Mouth_Slightly_Open | 32 Smiling |
| 3 Attractive | 13 Bush_Eyebrows | 23 Mustache | 33 Straight_Hair |
| 4 Bags_Under_Eyes | 14 Chubby | 24 Narrow_Eyes | 34 Wavy_Hair |
| 5 Bald | 15 Double_Chin | 25 No_Beard | 35 Wearing_Earrings |
| 6 Bangs | 16 Eyeglasses | 26 Oval_Face | 36 Wearing_Hat |
| 7 Big_Lips | 17 Goatee | 27 Pale_Skin | 37 Wearing_Lipstick |
| 8 Big_Nose | 18 Gray_Hair | 28 Pointy_Nose | 38 Wearing_Necklace |
| 9 Black_Hair | 19 Heavy_Makeup | 29 Receding_Hairline | 39 Wearing_Necktie |
| 10 Blond_Hair | 20 High_Cheekbones | 30 Rosey_Cheeks | 40 Young |

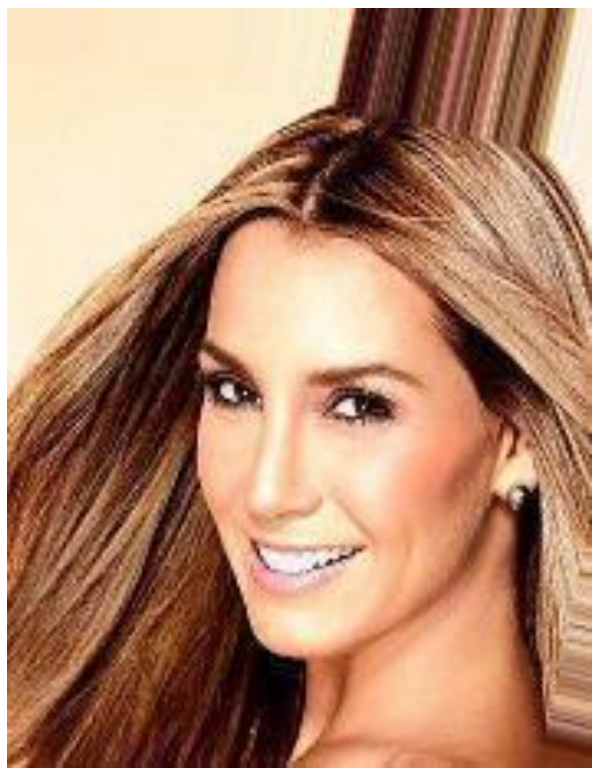

000001.jpg (visible image)

| | | | | | | | | | | | | | | | | | | | | | | | | | | | | | | | | | | | | | | | |
|---|---|---|---|---|---|---|---|---|---|---|---|---|---|---|---|---|---|---|---|---|---|---|---|---|---|---|---|---|---|---|---|---|---|---|---|---|---|---|---|
| 1 | 2 | 3 | 4 | 5 | 6 | 7 | 8 | 9 | 10 | 11 | 12 | 13 | 14 | 15 | 16 | 27 | 18 | 19 | 20 | 21 | 22 | 23 | 24 | 25 | 26 | 27 | 27 | 28 | 30 | 31 | 32 | 33 | 34 | 35 | 36 | 37 | 38 | 39 | 40 |

000001.jpg (binarized feature representation; -1 feature absent (red). 1 feature present (green))

| | | | | | | | | | | | | | | | | | | | | | | | | | | | | | | | | | | | | | | | |
|---|---|---|---|---|---|---|---|---|---|---|---|---|---|---|---|---|---|---|---|---|---|---|---|---|---|---|---|---|---|---|---|---|---|---|---|---|---|---|---|
| -1 | 1 | 1 | -1 | -1 | -1 | -1 | -1 | -1 | -1 | -1 | 1 | -1 | -1 | -1 | -1 | -1 | -1 | 1 | 1 | -1 | 1 | -1 | -1 | -1 | -1 | -1 | 1 | -1 | -1 | -1 | 1 | 1 | -1 | 1 | -1 | 1 | -1 | -1 | 1 |

Figure 8.
Representation of features in image 000001, (178 x 218 pixels), a female labeled as attractive from the *CelebA* dataset.

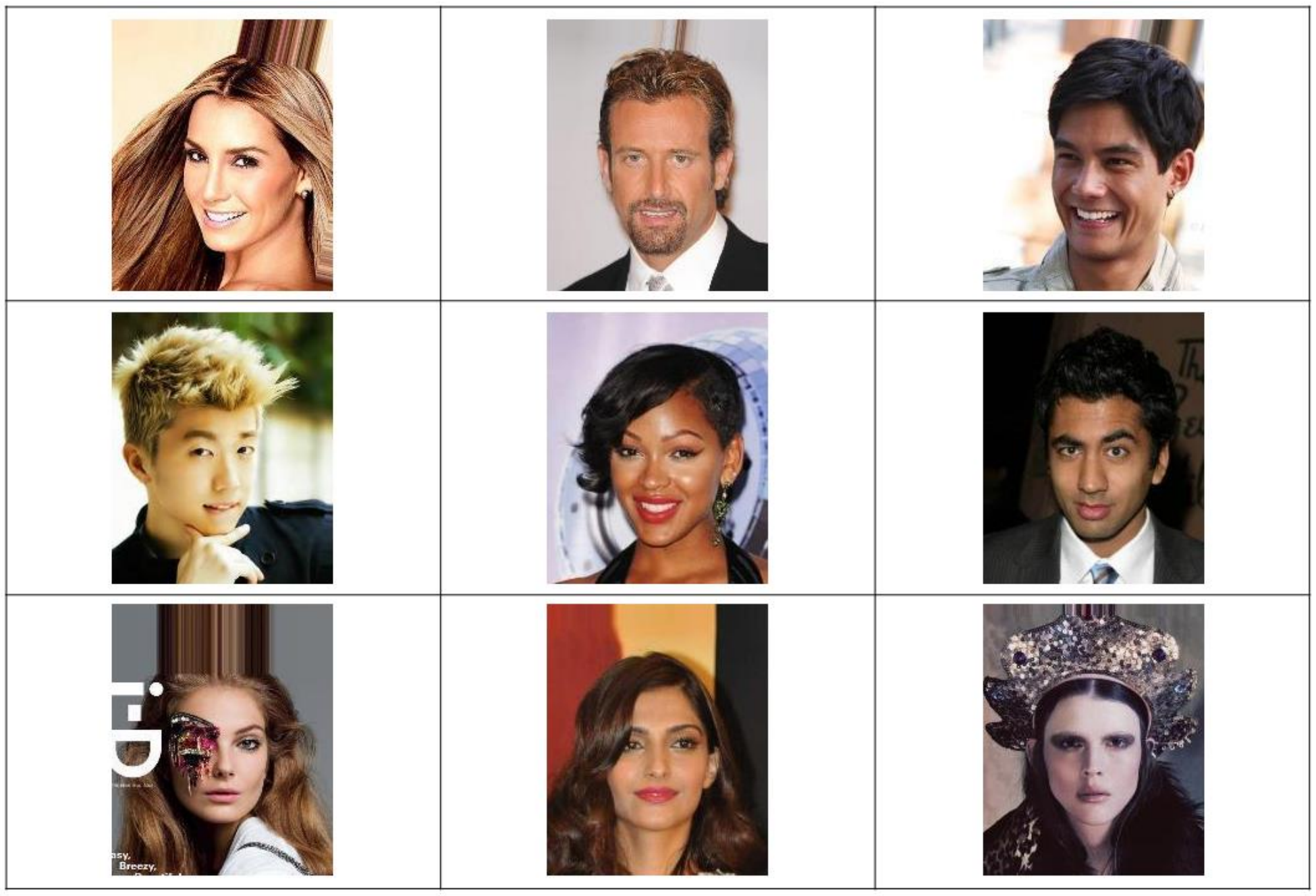

Figure 9.
Examples of the attribute ‘attractive’ from the *CelebA* dataset

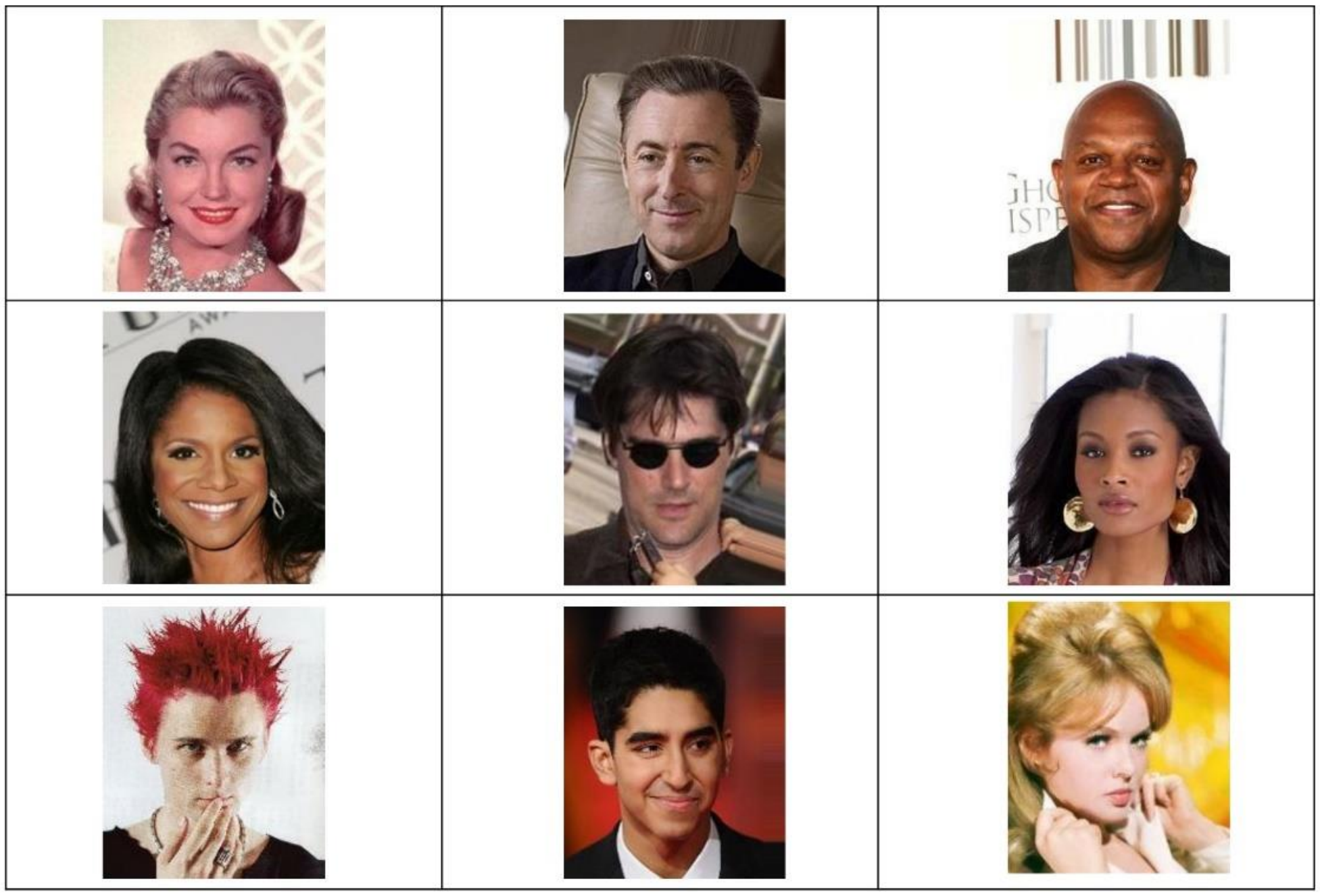

Figure 10.
Examples of the attribute ‘unattractive’ from the *CelebA* dataset

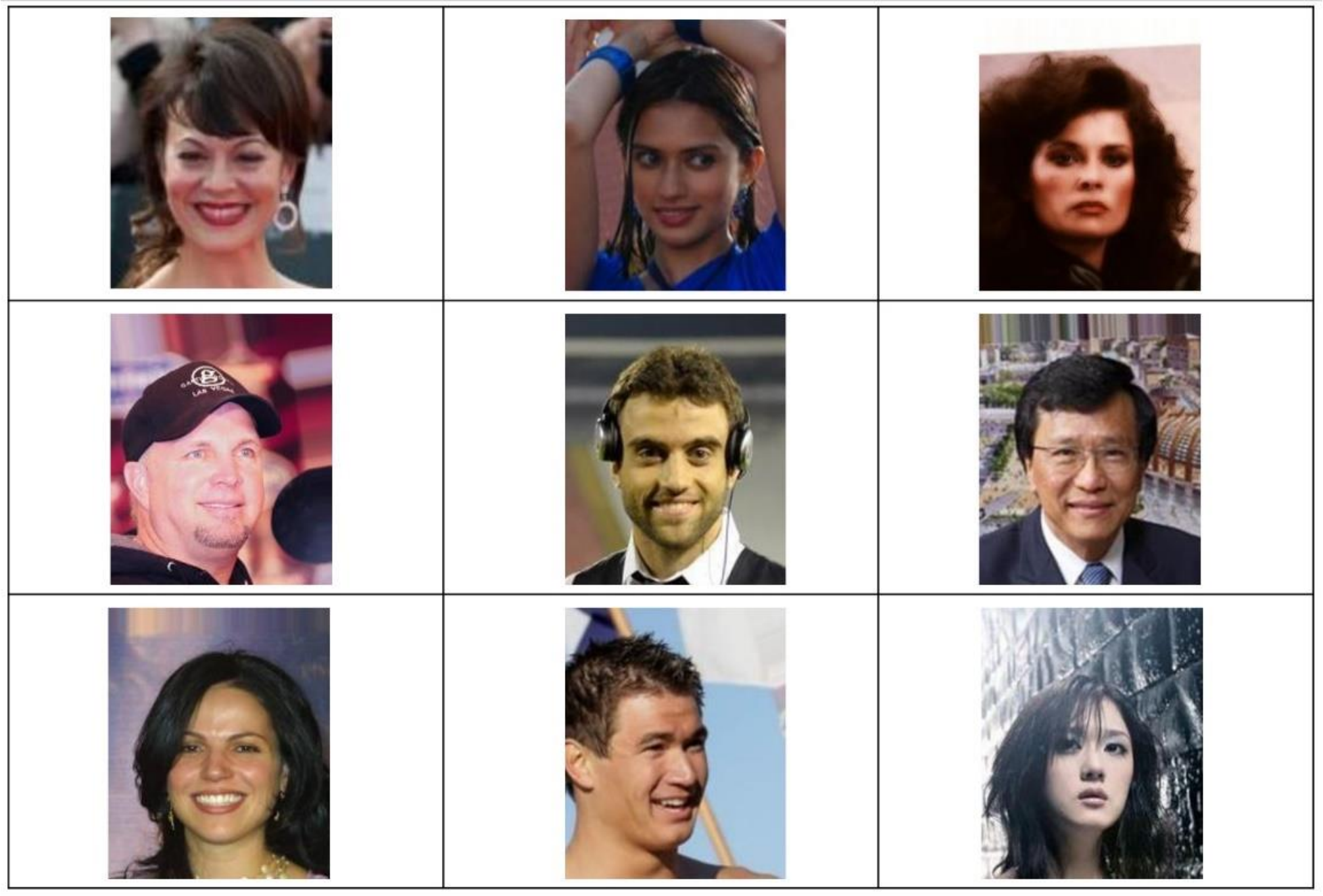

Figure 11.
Confusion matrix. Attractive predicted as unattractive

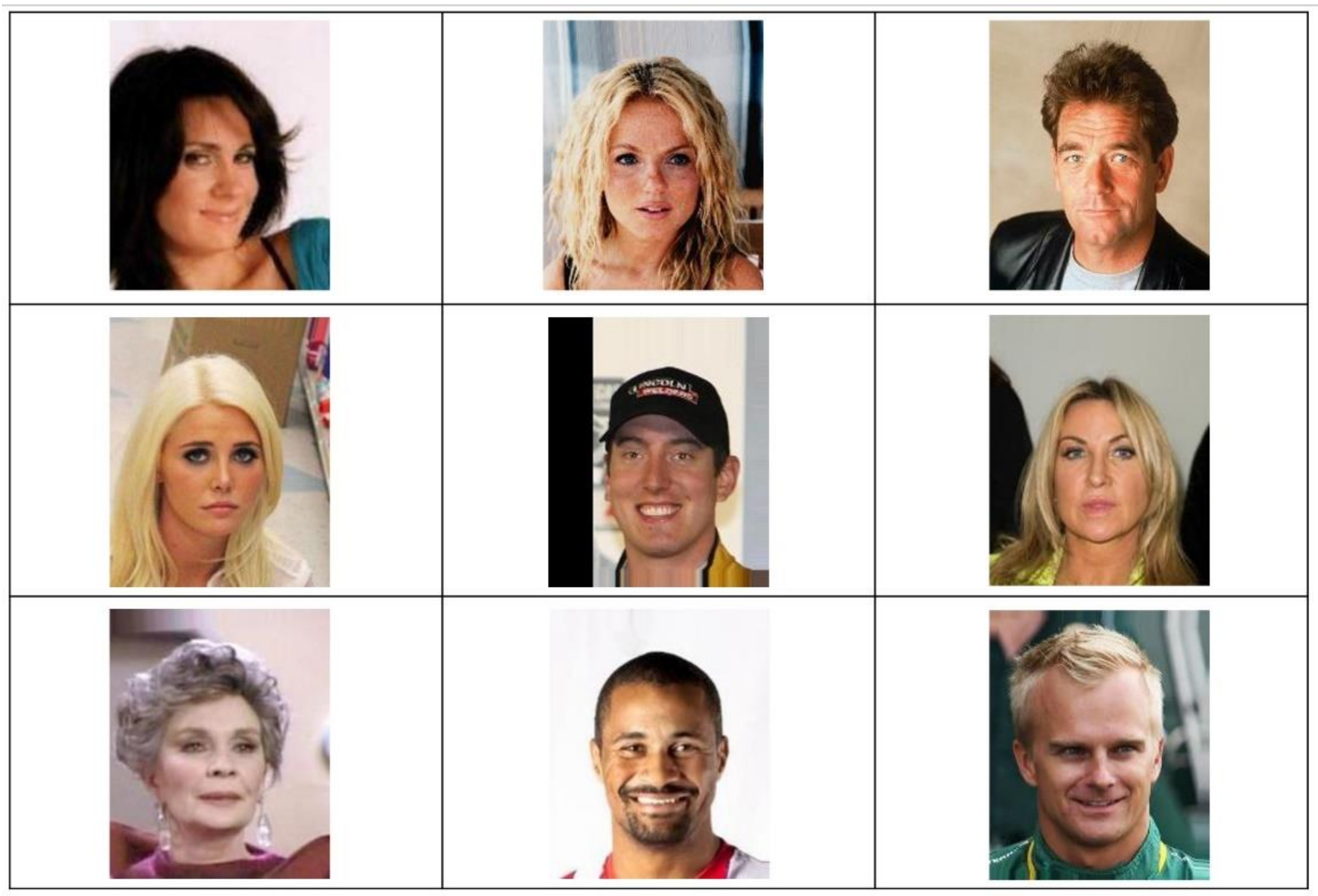

Figure 12.
Confusion matrix. Unattractive predicted as attractive

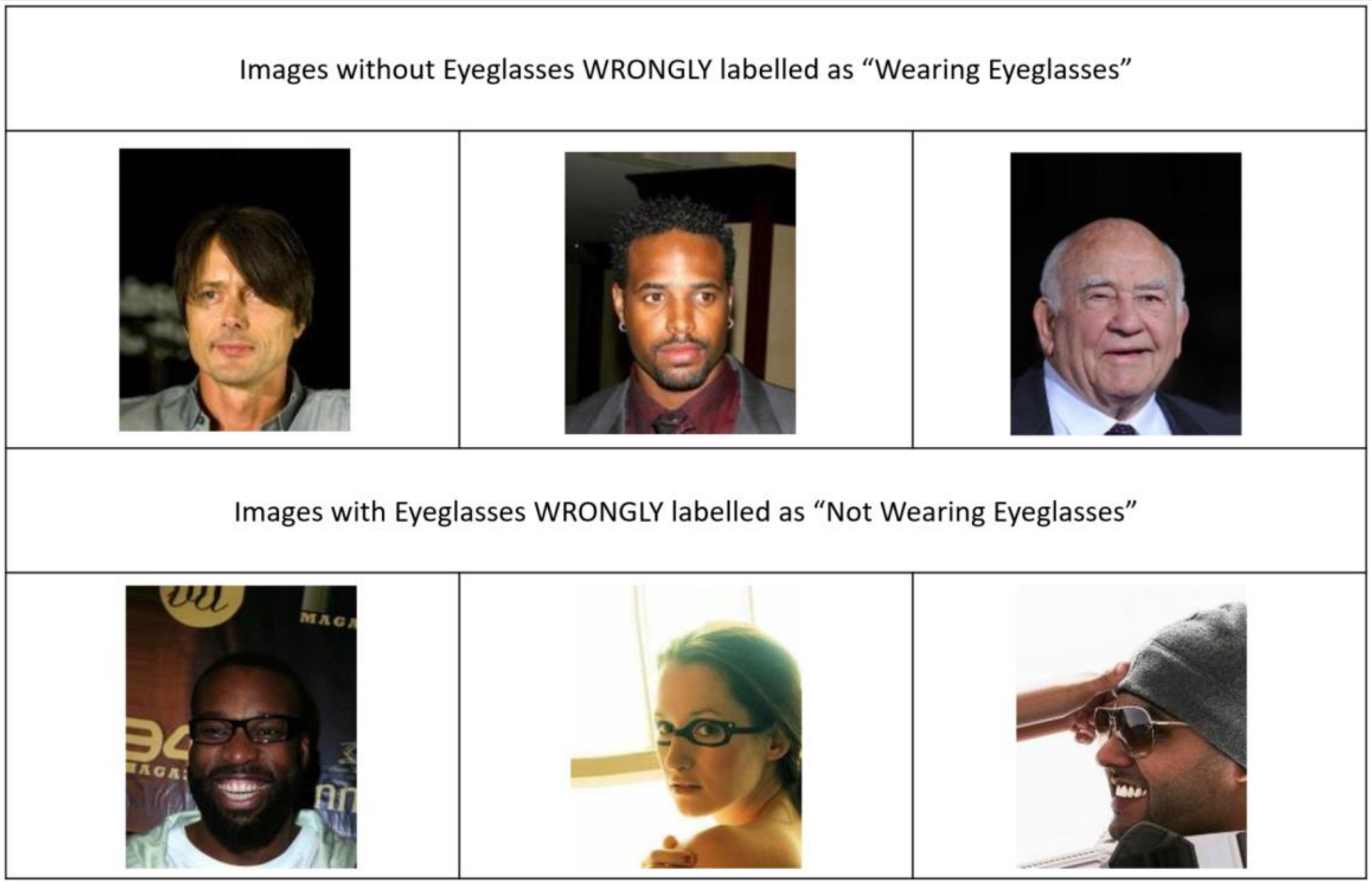

Figure 13.
Mislabeled data examples in the eyeglasses category.

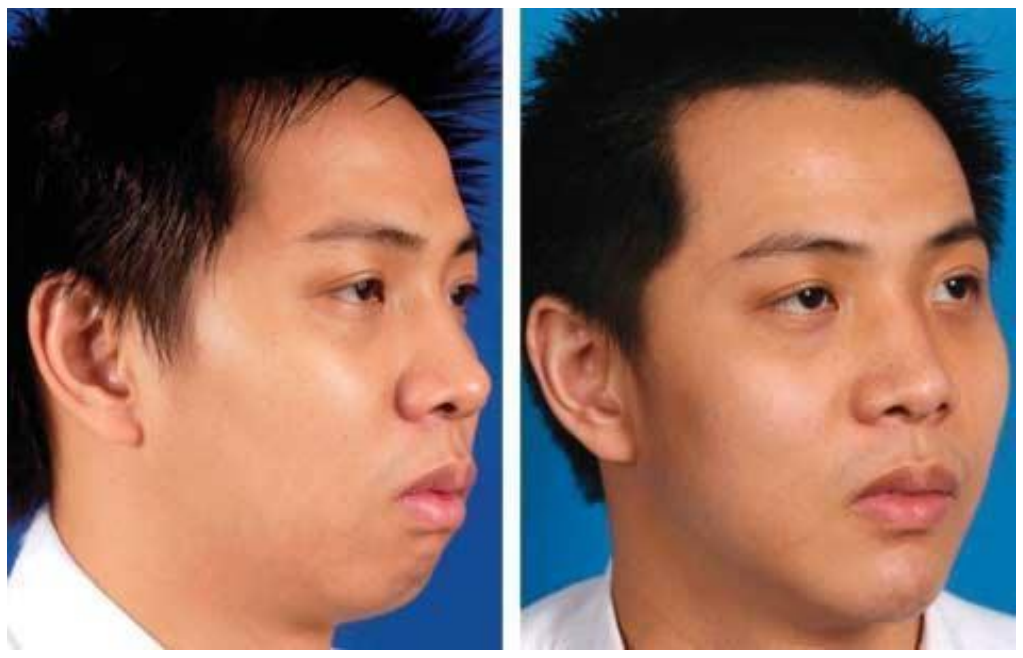

Figure 14.
Retrognathia, or “poor man’s chin”, before (left) and after (right) sliding genioplasty. The author, a plastic surgeon, reports that the patient launched a successful real estate career 6 months after the surgical intervention [Wong 2010].